\documentclass[%
 aip,
 amsmath,amssymb,
 reprint,%
 groupedaddress, %
]{revtex4-1}

\usepackage{graphicx}
\usepackage{dcolumn}
\usepackage{bm}

\usepackage[utf8]{inputenc}
\usepackage[T1]{fontenc}
\usepackage{mathptmx}
\usepackage{etoolbox}
\usepackage{enumerate}

\makeatletter
\def\@email#1#2{%
 \endgroup
 \patchcmd{\titleblock@produce}
  {\frontmatter@RRAPformat}
  {\frontmatter@RRAPformat{\produce@RRAP{*#1\href{mailto:#2}{#2}}}\frontmatter@RRAPformat}
  {}{}
}%
\makeatother
\begin{document}

\preprint{AIP/123-QED}

\title{SiGe/Si(111)/SiGe heterostructure for Si spin qubits with electrons confined in L valley of conduction band}
\author{Takafumi Tokunaga}
\email{t.tokunaga@ruri.waseda.jp}
\author{Hiromichi Nakazato}%
\affiliation{Department of Physics, Waseda University, Tokyo 169-8555, Japan.}%

\date{\today}

\begin{abstract}
ABSTRACT

\noindent
In Si(111) crystals, a strong biaxial tensile strain applied within the (111) plane is considered to shift the lowest energy point of the conduction band from the $\Delta$ valley to the L valley. Electrons confined in this L valley experience a splitting of their quadruply degenerate energy levels into an undegenerate single-level ground state (L1) and a triply degenerate excited state (L3). The energy of the single-level ground state is sufficiently low relative to the energies of the L3 valley and the $\Delta$ valley, making it optimal as a two-level system for a qubit. Using deformation potential theory and incorporating quantum effects from electron confinement in the SiGe/Si(111)/SiGe structure, we determine the value of the biaxial tensile strain causing the shift of the conduction band energy minimum from the $\Delta$ valley to the L valley, along with the corresponding Ge concentration. We also calculate the critical thickness for the plastic relaxation of the Si quantum well under this large biaxial tensile strain and examine the feasibility of realizing it as a SiGe/Si(111)/SiGe heterostructure. 
\end{abstract}

\maketitle

\section{\label{sec:level1}Introduction}

Electron-spin-qubit devices based on Si are being vigorously investigated due to their excellent compatibility with Si semiconductor technology, as exemplified by CMOS\cite{maurand2016cmos,camenzind2022hole,klemt2023electrical,geyer2024anisotropic,steinacker2025industry}. In Si(001) crystals used in CMOS, the lowest energy point of the conduction band in its band structure lies on the $X_{0} \text{ point on the} $ $\Gamma - \Delta$ axis and is sixfold degenerate in the bulk state. Henceforth, this point shall be referred to as the $\Delta$ valley.

When the SiGe/Si/SiGe layer is continuously epitaxially grown on Si(001) crystal, and biaxial tensile strain is applied to the Si layer sandwiched within this SiGe layer, the sixfold degeneracy lifts. The system splits into an excited state with fourfold degeneracy and a ground state with twofold degeneracy\cite{friesen2007valley,saraiva2009physical,saraiva2011intervalley}. This doubly degenerate ground state exhibits significant quantum effects as a result of the thin film nature of the sandwiched Si layer with thicknesses ranging from several to tens of nanometers. Consequently, even variations in the thickness of the atomic-level cause the degeneracy to lift, inducing valley splitting\cite{paquelet2022atomic,losert2023practical,lima2023interface}.

This valley splitting exhibits an energy difference between the two levels that varies from 20 $\mu$eV to 300 $\mu$eV \cite{borselli2011measurement,neyens2018critical,hollmann2020large,degli2024low}; this instability is termed as pseudo-degeneracy. The two-level system of qubits relies on an energy difference of approximately tens of $\mu$eV, depending on the spin orientation due to Zeeman splitting. A pseudo-degeneracy exists in the ground state; if this degeneracy is lifted, it causes disruption to the two-level system.\cite{hollmann2020large,yang2013spin,borjans2019single}. 

In the integration of silicon-based electron-spin-qubit devices, the unstable splitting of this doubly degenerate ground state presents a major challenge\cite{losert2024strategies,david2024long}. Various attempts have been made to significantly increase the energy difference during valley splitting, making it much larger than the Zeeman splitting energy difference, but a solution has not yet been achieved\cite{losert2023practical,feng2022enhanced,woods2024coupling,mcjunkin2022sige}.

As a possible solution to the above difficulty, this research investigates a qubit device that confines an electron in the L valley, by shifting the lowest conduction band energy point from the $\Delta$ valley to the L valley. To shift the lowest point of the conduction band energy to the L valley, Si(111) crystals are used. Specifically, $\text{Si}_{1-x}\text{Ge}_{x}$/Si(111)/$\text{Si}_{1-x}\text{Ge}_{x}$ (${x}\geqq0.94$) – a SiGe layer with a high Ge composition ratio – or a pure Ge layer sandwiching the Si layer under a large biaxial tensile strain.

In this structure, the lowest point of the conduction band energy is in the L valley. The fourfold degeneracy splits into an excited state L3 with threefold degeneracy and a ground state L1 with no degeneracy, due to the biaxial tensile strain. This ground state L1 exhibits no unstable pseudo-degeneracy. Furthermore, the $\Delta$ valley remains sixfold degenerate and can be positioned at an energy level approximately 70 meV higher, ensuring that it poses no hindrance whatsoever to the two-level system.

The reasons for using Si(111) crystals are as follows: compared to Si(001) or Si(011) crystals, the magnitude of the biaxial tensile strain required to position the lowest point of the conduction band energy in the L valley can be reduced; and the high symmetry of the L valley in Si(111) crystals facilitates the reduction of spin-orbit interaction and is advantageous for maintaining coherent states\cite{tokunaga2025111}.

The change in energy at the conduction band's $\Delta$ and L valleys due to biaxial tensile strain is calculated using the theory of deformation potential\cite{bardeen1950deformation,herring1956transport}. Within the range of small strain, the energy change varies linearly with strain. However, as the strain increases, the nonlinearity becomes significant and cannot be ignored; therefore, this is included in the analysis.

Furthermore, in the structure $\text{Si}_{1-x}\text{Ge}_{x}/ \text{Si(111)}/ \text{Si}_{1-x}\text{Ge}_{x}$ (${x}\geqq0.94$), when the thickness of the interposed Si layer reaches approximately several nanometers, the quantum effects become significant. This leads to differences in the energy change due to variations in the effective mass at each valley within the band structure. In calculating quantum effects, the energy levels of electrons within a well-type potential—where the offset of the conduction band's lowest energy point at the epitaxial growth film interface is treated as the barrier height—are computed quantum mechanically.

In Chapter II of this paper, we first summarize the energy changes due to biaxial tensile strain using deformation potential theory, then the energy changes due to quantum effects, and finally the combined effect integrating these biaxial tensile strain and quantum effects. When a substantial tensile strain is imposed on the Si(111) crystal and the energy caused by this enormous strain increases with film thickness, once the energy to generate dislocations exceeds the threshold, the film relaxes and ceases to be a strained Si film. We shall also discuss this critical film thickness. Furthermore, we briefly touch on the challenges associated with the epitaxial growth technique. Chapter III presents the discussion and conclusions.

\section{Theoretical calculations of energy shifts in the L and $\Delta$ valleys}

\subsection{\label{sec:level2}The effect of strain on energy}

The deformation potential theory was proposed by Bardeen and Shockley and was later systematized by Herring, Vogt, and Van de Walle\cite{bardeen1950deformation,herring1956transport,van1985theoretical,van1986theoretical}. This theory provides a framework for deriving, through relatively straightforward calculations, how lattice strain alters the electronic energy band structure using deformation potentials obtained via self-consistent density functional calculations. In $\text{Si}_{1-x}\text{Ge}_{x}/\text{Si}$ structures, the research of Rieger and Vogl has been particularly influential\cite{rieger1993electronic}. The work by Hinckley and Singh discusses how the band structure in the $\text{Si}_{1-x}\text{Ge}_{x}/\text{Si}$ structure varies depending on the orientation of the Si crystal\cite{hinckley1990influence}.

An epitaxial film with a lattice constant different from its underlying substrate experiences a uniform biaxial tensile strain $\epsilon_{\|}>0$ or a biaxial compressive strain $\epsilon_{\|}<0$ within the plane of the film. This biaxial strain in the plane $\epsilon_{\|}$ is defined by the bulk lattice constants $a_{S}$ and $a_{L}$ of the underlying substrate material and the epitaxial growth film layer material, respectively, as follows:
\begin{equation}
\epsilon_{\|} = \frac{a_{S}}{a_{L}}-1.
\end{equation}

It is assumed that the underlying substrate is sufficiently thick to retain its bulk lattice spacing, and that the epitaxial growth film perfectly matches the substrate's lattice spacing. This paper focuses on Si(111) crystals, and their coordinate system is denoted as $x^{\prime},y^{\prime},z^{\prime}$. The principal crystal axes (100), (010), (001) of the Si(001) crystal are related to the $x, y, z$ coordinate system by the rotation $U$ defined below\cite{hinckley1990influence}
\begin{equation}
\begin{aligned}
& U=\left[\begin{array}{ccc}
\cos \varphi \cos \theta & -\sin \varphi & \cos \varphi \sin \theta \\
\sin \varphi \cos \theta & \cos \varphi & \sin \varphi \sin \theta \\
-\sin \theta & 0 & \cos \theta
\end{array}\right], \\
& U^{(111)}=\left[\begin{array}{ccc}
\frac{1}{\sqrt{6}} & \frac{-1}{\sqrt{2}} & \frac{1}{\sqrt{3}} \\
\frac{1}{\sqrt{6}} & \frac{1}{\sqrt{2}} & \frac{1}{\sqrt{3}} \\
-\sqrt{\frac{2}{3}} & 0 & \frac{1}{\sqrt{3}}
\end{array}\right].
\end{aligned}
\end{equation}

The in-plane strain tensor is expressed as
\begin{equation}
\overleftrightarrow{\varepsilon^{\prime}}=\left(\begin{array}{ccc}
\varepsilon_{\|} & 0 & \varepsilon_{13}^{\prime} \\
0 & \varepsilon_{\|} & \varepsilon_{23}^{\prime} \\
\varepsilon_{31}^{\prime} & \varepsilon_{32}^{\prime} & \varepsilon_{33}^{\prime}
\end{array}\right),
\end{equation}
where
\begin{equation}
\varepsilon_{11}^{\prime}=\varepsilon_{22}^{\prime}=\varepsilon_{\|}, \quad \varepsilon_{12}^{\prime}=\varepsilon_{21}^{\prime}=0.
\end{equation}

As the substrate applies uniform in-plane stress to the epitaxial growth film, let the stress tensor be denoted by $\sigma^{\prime}_{ij}$,

\begin{equation}
\sigma_{33}^{\prime}=\sigma_{23}^{\prime}=\sigma_{13}^{\prime}=0,
\end{equation}
while $\sigma^{\prime}_{11}, \sigma^{\prime}_{22}$ and $\sigma^{\prime}_{12}$ become non-zero. From Hooke's law, using the elastic stiffness tensor $C^{\prime}_{ijkl}$,
 it can be expressed as $\sigma_{i j}^{\prime}=\sum_{k,l=1}^3 C_{i j k l}^{\prime} \varepsilon_{k l}^{\prime}$ and therefore
\begin{equation}
\sum_{i,j=1}^3C_{33 i j}^{\prime} \varepsilon_{i j}^{\prime}=\sum_{i,j=1}^3C_{23 i j}^{\prime} \varepsilon_{i j}^{\prime}=\sum_{i,j=1}^3C_{13 i j}^{\prime} \varepsilon_{i j}^{\prime}=0.
\end{equation}
Since Si(001) crystals and Si(111) crystals possess the characteristic that their normal is an axis of rotation with ${n}$-fold symmetry $(n>2)$,
\begin{equation}
\varepsilon_{13}^{\prime}=\varepsilon_{31}^{\prime}=\varepsilon_{23}^{\prime}=\varepsilon_{32}^{\prime}=0.
\end{equation}
Therefore,  if we set $\varepsilon_{33}^{\prime}=\varepsilon_{\perp}$,
\begin{equation}
\varepsilon_{33}^{\prime}=\varepsilon_{\perp}=-\frac{C_{3311}^{\prime} \varepsilon_{11}^{\prime}+C_{3322}^{\prime} \varepsilon_{22}^{\prime}}{C_{3333}^{\prime}}=-\left(\frac{C_{3311}^{\prime}+C_{3322}^{\prime}}{C_{3333}^{\prime}}\right) \varepsilon_{\|},
\end{equation}
\begin{equation}
C_{33 k k}^{\prime}=\sum_{\alpha, \beta, i, j=1}^3 U_{\alpha 3} U_{\beta 3} U_{i k} U_{j k} C_{\alpha \beta i j}.
\end{equation}
Since the elastic stiffness tensor for crystals with $O_{h}$ symmetry, such as SiGe, only possesses $C_{11},C_{12}$ and $C_{44}$ components, we can express as:
\begin{equation}
\varepsilon_{\perp}=-\frac{\frac{2}{3} C_{11}+\frac{4}{3} C_{12}-\frac{4}{3} C_{44}}{\frac{1}{3} C_{11}+\frac{2}{3} C_{12}+\frac{4}{3} C_{44}} \varepsilon_{\|}=-\frac{2 C_{11}+4 C_{12}-4 C_{44}}{C_{11}+2 C_{12}+4 C_{44}} \varepsilon_{\|}.
\end{equation}
From the above, the biaxial tensile strain tensor in the Si(111) coordinate system is expressed as
\begin{equation}
\overleftrightarrow{\varepsilon^{\prime}}=\left(\begin{array}{ccc}
\varepsilon_{\|} & 0 & 0 \\
0 & \varepsilon_{\|} & 0 \\
0 & 0 & \varepsilon_{\perp}
\end{array}\right).
\end{equation}
Transformation into a coordinate system aligned with the Si(001) crystal axis can be expressed as
\begin{equation}
\overleftrightarrow{\varepsilon}=\left(\begin{array}{ccc}
\frac{2 \varepsilon_{\|}+\varepsilon_{\perp}}{3} & \frac{\varepsilon_{\perp}-\varepsilon_{\|}}{3} & \frac{\varepsilon_{\perp}-\varepsilon_{\|}}{3} \\
\frac{\varepsilon_{\perp}-\varepsilon_{\|}}{3} & \frac{2 \varepsilon_{\|}+\varepsilon_{\perp}}{3} & \frac{\varepsilon_{\perp}-\varepsilon_{\|}}{3} \\
\frac{\varepsilon_{\perp}-\varepsilon_{\|}}{3} & \frac{\varepsilon_{\perp}-\varepsilon_{\|}}{3} & \frac{2 \varepsilon_{\|}+\varepsilon_{\perp}}{3}
\end{array}\right).
\end{equation}
Using this $\overleftrightarrow{\varepsilon}$, we investigate the change in energy in each valley $\Delta$ or L due to biaxial tensile stress, employing the theory of the deformation potential.
\begin{table*}
\caption{\label{tab:table1}The values of the deformation potentials in units of eV and their sources. Priority is given to papers in which the values of $\Xi_{d}^{(\Delta)}$, $\Xi_{u}^{(\Delta)}$, $\Xi_{d}^{(L)}$ and $\Xi_{u}^{(L)}$ are all provided.}
\begin{ruledtabular}
\begin{tabular}{c|cccccccccccccc|ccc}
Deformation&\multicolumn{14}{c}{Theory}&\multicolumn{3}{c}{Experiment}\\
potential[eV]&Ref.\cite{chelikowsky1976nonlocal}&Ref.\cite{van1986theoretical}&Ref.\cite{friedel1989local}&Ref.\cite{schmid1990calculated}&Ref.\cite{hinckley1990influence}&Ref.\cite{tserbak1993unified}&Ref.\cite{rieger1993electronic}&Ref.\cite{yoder1994first}&Ref.\cite{fischetti1996band}&Ref.\cite{rideau2006strained}&Ref.\cite{ungersboeck2007effect}&Ref.\cite{li2021deformation}&Ref.\cite{yang2024uncovering}&Ref.\cite{williams2025improved}&Ref.\cite{laude1971effects}&Ref.\cite{li1991shear}&Ref\cite{chen2011conduction}.
\\ \hline
$\Xi_{u}^{(\Delta)}$&9.0&9.16&8.47&8.0&9.2&8.86&9.29&   &10.5&9.01\footnotemark[2]&9.29&8.84&9.0&10.03\footnotemark[2]&8.6&11.1&9.1\\
$\Xi_{d}^{(\Delta)}$\footnotemark[1]&&1.10&1.03&0.63&&-1.62&-0.71&1.2&1.1&0.94\footnotemark[2]&1.1&1.01&0.8&1.47\footnotemark[2]&0.73&&\\
$\Xi_{u}^{(L)}$&15.9&16.14&12.35&&&&&&18.0&15.1\footnotemark[2]&&&&&&&18.1\\
$\Xi_{d}^{(L)}$\footnotemark[1]&&-6.00&-4.90&&&&&&-7.0&-6.06\footnotemark[2]&&&&&&&\\ 
\end{tabular}
\end{ruledtabular}
\footnotetext[1]{$\Xi_{d}$ is obtained in the form of the hydrostatic deformation potential $\Xi_d+\Xi_u / 3-a$ ($a$: the hydrostatic deformation potential at the top of the valence band $\Gamma$).}
\footnotetext[2]{Due to differences in calculation methods and fitting techniques, multiple values are obtained.}
\end{table*}

Let the first-order change in energy in each valley ${\alpha}$ be indicated by $\Delta E_{1}^{(\alpha)}$, the dilatational deformation potential by $\Xi_{d}^{(\alpha)}$, and the uniaxial deformation potential by $\Xi_{u}^{(\alpha)}$. $\Delta E_{1}^{(\alpha)}$ is expressed as\cite{van1989band} 
\begin{equation}
\Delta E_{1}^{(\alpha)}=\left(\Xi_d^{(\alpha)} \overleftrightarrow{1}+\Xi_u^{(\alpha)}\left\{\overrightarrow{\mathbf{a}}_\alpha \overrightarrow{\mathbf{a}}_\alpha\right\}\right): \overleftrightarrow{\varepsilon},
\end{equation}
where $\overleftrightarrow{1}$ is the unit tensor, $\overrightarrow{\mathbf{a}}_\alpha$ is a unit vector parallel to the $\overrightarrow{k}$ (the wave vector defined within the Brillouin zone) of the valley $\alpha$, :(colon) denotes a double-dotted product and $\{ \}$ denotes a dyadic product.

The $\Delta$ valley remains sixfold degenerate under strain and is therefore denoted as $\Delta 6$. The L valley splits its fourfold degeneracy into a single non-degenerate state and a threefold degenerate state, denoted L1 and L3, respectively. The first-order linear terms $\Delta {E}_1$ of the energy changes for L1, L3, and $\Delta 6$ are expressed respectively,
\begin{equation}
\begin{aligned}
\Delta E_1^{(L 1)} & =\Xi_d^{(L)}\left(2 \varepsilon_{\|}+\varepsilon_{\perp}\right)+\Xi_u^{(L)} \varepsilon_{\perp}, \\
\Delta E_1^{(L 3)} & =\Xi_d^{(L)}\left(2 \varepsilon_{\|}+\varepsilon_{\perp}\right)+\frac{1}{9} \Xi_u^{(L)}\left(8 \varepsilon_{\|}+\varepsilon_{\perp}\right), \\
\Delta E_1^{(\Delta 6)} & =\Xi_d^{(\Delta)}\left(2 \varepsilon_{\|}+\varepsilon_{\perp}\right)+\frac{1}{3} \Xi_u^{(\Delta)}\left(2 \varepsilon_{\|}+\varepsilon_{\perp}\right) .
\end{aligned}
\end{equation}
Note that the specific values of the elastic constants $C_{11}=165.7 \mathrm{GPa},\  C_{12}=63.9 \mathrm{GPa},\  C_{44}=79.6 \mathrm{GPa}$ \cite{ungersboeck2007effect,LandoltBornstein2001:sm_lbs_978-3-540-31355-7_212}
 in Eq.(10) yields 
\begin{equation}
\begin{aligned}
\varepsilon_{\perp} \cong-0.44 \varepsilon_{\|}.
\end{aligned}
\end{equation}

Much research has been conducted on the deformation potentials, $\Xi_{d}$ and $\Xi_{u}$, and the specific numerical values are summarized in Table I. In this Table, $\Xi_{u}$ is obtained directly and with high precision from uniaxial stress tests such as cyclotron resonance and piezoresistance measurements. On the other hand, $\Xi_{d}$ is obtained indirectly by applying hydrostatic pressure, in the form of the hydrostatic deformation potential $\Xi_{d} + \Xi_{u}/3 -a$ \ ($a$: the hydrostatic deformation potential at the top of the valence band $\Gamma$) and is therefore slightly less accurate.

In this paper, we use the values of Van de Walle’s study\cite{van1986theoretical}, for which the values of $\Xi_{d}^{(\Delta)},\  \Xi_{u}^{(\Delta)},\  \Xi_{d}^{(L)},\  \Xi_{u}^{(L)}$ are all available and have been used as standard in subsequent research. However, since changes in the values of these deformation potentials have a significant impact on the conclusions of this study, we will introduce a variation of approximately 10$\%$ in these values in the section "Discussion and Conclusion" to investigate the resulting effects.
We therefore take the values of the deformation potentials to be $\Xi_u^{(\Delta)}=9.16 \mathrm{eV}$, $\Xi_d^{(\Delta)}=1.10 \mathrm{eV}$, $\Xi_u^{(L)}=16.14 \mathrm{eV}$, $\Xi_d^{(L)}=-6.00 \mathrm{eV}$ in Eq.(14), and we obtain the following expressions for the first-order linear term $\Delta E_{1}$ of the energy changes

\begin{equation}
\begin{aligned}
&\Delta E_1^{(L 1)} \cong-16.46 \varepsilon_{\|} \ \text{eV}, \ \Delta E_1^{(L 3)} \cong 4.20 \varepsilon_{\|} \ \text{eV}, \\ 
& \Delta E_1^{(\Delta 6)} \cong 6.48 \varepsilon_{\|} \  \text{eV}.
\end{aligned}
\end{equation}
Research on the non-linear effects of the deformation potentials in Si crystals has evolved into studies aimed at increasing the speed of CMOS devices by applying greater strain to Si\cite{fischetti1996band}. Much of this research uses first-principle calculations, such as DFT, to investigate band structures in regions of high strain where experimental results are difficult to obtain\cite{van1985theoretical}. 

As strain increases, as studied by Nielsen $\textit{et al.}$, a quantity representing changes in internal degrees of freedom beyond what can be described solely by changes in the unit cell's external shape becomes necessary. This quantity is termed the internal strain parameter $\zeta$ (also known as the Kleinman parameter)\cite{kleinman1962deformation,kleinman1963deformation,goroff1963deformation,li1991shear,nielsen1985stresses}.

The Si crystal has a lattice structure in which each unit cell contains two atoms. When the lattice shape changes due to strain, the unit cell vectors (axes a, b and c) change for small strains; however, as the strain increases, the relative positions of the two atoms within the unit cell also change. The relative positions of these two atoms within the unit cell constitute an internal degree of freedom, and $\zeta$ represents the change in these positions.

Since $\zeta$ itself changes due to strain, when incorporating the second-order effect in the form $\Delta E_2^{(\alpha)}=\sum D_{i j k l}^{(\alpha)} \varepsilon_{i j} \varepsilon_{k l}$,  the fitting parameter $D_{ijkl}^{(\alpha)}$ is determined using first-principle calculations\cite{rieger1993electronic,fischetti1996band,li2021deformation}. Consequently, the value of $D_{i j k l}^{(\alpha)}$ used in the first-principle calculations depends on the parameter settings of the pseudopotential and the $\text{k$\cdot$p}$ method. Therefore, in this study, taking into account a certain degree of variation, we first use its approximate average to calculate the second-order nonlinear terms $\Delta E_{2}$ of the changes in each valley as follows:
\begin{equation}
\begin{aligned}
& \Delta E_{2}^{(L1)} = -22.5\epsilon_{\|}^2\ \text{eV}, \ \Delta E_{2}^{(L3)} = -15.0\epsilon_{\|}^2\ \text{eV}, \\
& \Delta E_{2}^{(\Delta6)} = -10.0\epsilon_{\|}^2 \ \text{eV}.
\end{aligned}
\end{equation}
As the effect of variations in the value of $D_{i j k l}^{(\alpha)}$ also influences the conclusions of this study, we will conduct an analysis taking these variations into account in the section "Discussion and Conclusion".

Organize the energy at each valley of the band in the form 
\begin{equation}
E^{(\alpha)}(\varepsilon)=E_0^{(\alpha)}+\Delta E_1^{(\alpha)}+\Delta E_2^{(\alpha)},
\end{equation}
we obtain explicitly
\begin{equation}
\begin{aligned}
& E^{(L 1)}=2.10-16.46 \varepsilon_{\|}-22.5 \epsilon_{\|}^2 \ \text{eV}, \\
& E^{(L 3)}=2.10+4.20 \varepsilon_{\|}-15.0 \epsilon_{\|}^2 \ \text{eV}, \\
& E^{(\Delta 6)}=1.17+6.48 \varepsilon_{\|}-10.0 \epsilon_{\|}^2 \ \text{eV},
\end{aligned}
\end{equation}
where the energy values without strain $\text{at}\  0\ \mathrm{K}$\cite{macfarlane1958fine,bludau1974temperature,chelikowsky1976nonlocal,richard2004energy} have been set as ${E_{0}^{(L 1)}}={E_{0}^{(L 3)}}=2.10\ \mathrm{eV},\ {E_{0}^{(\Delta 6)}}=1.17\ \mathrm{eV}$.
\subsection{\label{sec:level2}The quantum effects arising from confinement in a well-type potential}

In the $\text{Si}_{1-x}\text{Ge}_{x}$/ Si(111) / $\text{Si}_{1-x}\text{Ge}_{x}$ structure, the electrons are confined within the sandwiched Si layers. This results from selecting the above Ge concentration to form a well-type potential in the band structure.
When the thickness of the sandwiched Si layer (taken in the ${z}$- direction) is of the order of several nanometers, it becomes comparable to the spread of the electron wave function, leading to significant quantum effects\cite{chen2010quantum}.

The lateral dimensions (${x}$ and ${y}$) of the device are of the order of tens to hundreds nanometers. Given the spread of the electron wave function, these dimensions are sufficiently large, which means that quantum effects due to confinement in the ${x}$ and ${y}$ directions can be neglected. Let the central position of this well-type potential be $z = 0$, the thickness of the sandwiched Si layer be ${t}$, and the offset amount of the lowest conduction band energy in the structure $\text{Si}_{1-x}\text{Ge}_{x}$ / Si(111) be $V_{0}$. That is, the height of the potential barrier is denoted as $V_{0}$.

With regard to the quantum effects of electrons confined within a thin film of Si, many studies use an infinite-height potential barrier for the sake of simplicity\cite{paul2004si,sverdlov2008two}. However, in order to achieve higher accuracy, we calculate quantum effects using the energy offset $V_{0}$ in the heterostructure as the height of the potential barrier.

The Schrödinger equation determining the energy of electrons confined within this well-type potential is as follows:
\begin{equation}
\begin{aligned}
& -\frac{\hbar^2}{2 m_{\mathrm{in}}^*} \frac{d^2 \psi(z)}{d z^2}=E_q^{(\alpha)} \psi(z), \quad-\frac{t}{2}<z<\frac{t}{2}, \\
& -\frac{\hbar^2}{2 m_{\mathrm{out}}^*} \frac{d^2 \psi(z)}{d z^2}+V_0 \psi(z)=E_q^{(\alpha)} \psi(z), \quad|z|>\frac{t}{2},
\end{aligned}
\end{equation}
where $m_\mathrm{in}^{*}$ and $m_\mathrm{out}^{*}$ denote the effective masses of the electron within and outside the well, respectively.

The ground state is given by an even function, so the wave function $\psi(z)$ reads as
\begin{equation}
\begin{aligned}
& \psi_{\text {in }}(z)=A \cos \left(k_{\text {in }} z\right), \quad k_{\text {in }}=\frac{\sqrt{2 m_{\text {in }}^* E_q^{(\alpha)}}}{\hbar}, \quad-\frac{t}{2}<z<\frac{t}{2}, \\
& \psi_{\text {out }}(z)=B e^{-k_{\text {out }}|z|}, \quad k_{\text {out }}=\frac{\sqrt{2 m_{\text {out }}^*\left(V_0-E_q^{(\alpha)}\right)}}{\hbar}, \quad|z|>\frac{t}{2} .
\end{aligned}
\end{equation}
From the continuity of the wave function and the continuity of the probability density (Ben Daniel-Duke boundary condition)\cite{bendaniel1966space}, we have
\begin{equation}
\begin{aligned}
A \cos \left(k_{\text {in }} \frac{t}{2}\right) & =B e^{-k_{\text {out }} \frac{t}{2}}, \\
-\frac{A k_{\text {in }}}{m_{\text {in }}^*} \sin \left(k_{\text {in }} \frac{t}{2}\right) & =-\frac{B k_{\text {out }}}{m_{\text {out }}^*} e^{-k_{\text {out }} \frac{t}{2}}.
\end{aligned}
\end{equation}
These two equations yield $\tan \left(k_{\text {in }} \frac{t}{2}\right)=\frac{m_{\text {in }}^* k_{\text {out }}}{m_{\text {out}}^* k_{\text {in }}}$, which is reduced to
\begin{equation}
\tan \left(\frac{\sqrt{2 m_\text{in}^* E_q^{(\alpha)}} t}{2 {\hbar}}\right)=\sqrt{\frac{m_\text{in}^*}{m_\text{out}^*} \frac{V_0-E_q^{(\alpha)}}{E_q^{(\alpha)}}}.
\end{equation}
We proceed to solve numerically for the eigenvalue $E_{q}^{(\alpha)}$.

As has been previously stated, the offset amount of the lowest point of the conduction band energy in the structure $\text{Si}_{1-x}\text{Ge}_{x}$ / Si(111) is denoted as $V_{0}$. The offset of the lowest conduction band energy has been quantified by Van de Walle $\textit{et al.}$, and is known to be 0.55 eV for the Ge / Si(001) / Ge structure and 0.28 eV for the Ge / Si(111) / Ge structure\cite{van1986theoretical}. Therefore, setting the height of the potential barrier $V_{0}$ at 0.28 eV, the quantum effect $E_q^{(\alpha)}$ is solved for each valley L1, L3 and $\Delta 6$ in the structure of the band, using the effective masses in Table II as parameters for the thickness of the film ${t}$ nm\cite{canali1975electron,tomizawa1993numerical}. 

Here, $m_\text{in}^{\star}$ and $m_\text{out}^{\star}$ are the effective masses in the ${z}$ direction in Si $m_{\text{Si}, z}^{\star}$ and SiGe $m_{\text{SiGe}, z}^{\star}$, respectively. Incidently, in the L1 valley, the effective mass $m_{\text{Si}, x,y}^{(L1)\star}$ within the (111) plane, i.e., in the ${x}$ and ${y}$ directions, becomes 0.12$m_{0}$\cite{rieger1993electronic,fischetti1996band}, making it very light. Consequently, it is expected that the mobility of electrons confined in the L1 valley within the (111) plane will be extremely high.
 
\begin{table}
\caption{\label{tab:table2}The effective masses in units of the electron mass in vacuum $m_{0}$ at each symmetry point of the band structure\cite{canali1975electron,tomizawa1993numerical}. $m_\text{in}^{\star}$ and $m_\text{out}^{\star}$ denote the effective masses within and outside the well-type potential, respectively.}
\begin{ruledtabular}
\begin{tabular}{c|ccc} 
symmetry point&$m_\text{in}^{\star}$\footnote{$m_\text{in}^{\star}$=$m_\text{Si,z}^{\star}$}&$m_\text{out}^{\star}$\footnote{$m_\text{out}^{\star}$=$m_\text{SiGe,z}^{\star}$}\\
\hline
L1 & 1.70 & 1.59&\\
L3 & 0.13 & 1.59&\\
$\Delta 6$ & 0.26 & 1.59&\\
\end{tabular}
\end{ruledtabular}
\end{table}

Figure 1 shows the graph of $E_q^{(\alpha)}$ for valleys L1, L3, and $\Delta6$, as a function of the thickness of the film ${t}$ nm. As can be seen in Fig.1, the quantum effects increase sharply when the film thickness ${t}$ reaches approximately 3 nm or less. Consequently, since the effective mass of $\Delta 6$ is lighter than that of L1, the rate of increase in energy is greater for $\Delta 6$ than for L1; therefore, the quantum effect causes the energy of L1 to approach that of $\Delta 6$.
\begin{figure}
\includegraphics[width=1.00\linewidth]{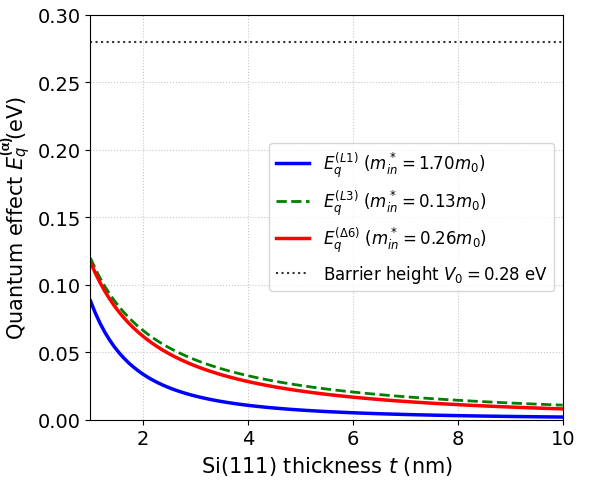}
\caption{\label{fig:epsart} Quantum effects $E_{q}^{(\alpha)}$ for valleys L1, L3, and $\Delta6$, as functions of Si(111) thickness ${t}$ nm. The height of the potential barrier $V_{0}$ is set as 0.28 eV. }
\end{figure}

\subsection{\label{sec:level2}Effects of Strain and Quantum Effects on Energy}
Integrating the quantum effect $E_q^{(\alpha)}$ obtained for each film thickness ${t}$ nm in the previous section with the effect due to strain, the energy in each valley L1, L3 and $\Delta6$ can be expressed in eV as follows:
\begin{equation}
\begin{aligned}
E^{(L l)} & =2.10-16.46 \varepsilon_{\|}-22.5 \varepsilon_{\|}^2+E_q^{(L 1)}\ \text{eV}, \\
E^{(L 3)} & =2.10+4.20 \varepsilon_{\|}-15.0 \varepsilon_{\|}^2+E_q^{(L 3)} \ \text{eV}, \\
E^{(\Delta 6)} & =1.17+6.48 \varepsilon_{\|}-10.0 \varepsilon_{\|}^2+E_q^{(\Delta 6)} \ \text{eV}.
\end{aligned}
\end{equation}
For example, for a film thickness of ${t}$ = 10 nm, the difference in the quantum effects $E_{q}^{(\alpha)}$ of L1, L3 and $\Delta 6$ is small, as shown in Fig.1. Therefore, their influence on the relative relationship of the total energy $E^{(\alpha)}$ is negligible. Figure 2 shows the total energy $E^{(\alpha)}$ at the thickness of the film ${t}$ = 10 nm, as the strain varies from 0\% to 5\%.
As can be seen in this figure, whilst $E^{(L1)}$ decreases rapidly as the tensile strain increases, $E^{(\Delta 6)}$ increases, resulting in an energy reversal. For a film thickness of ${t}$ = 10 nm, $E^{(L1)}$ becomes less than $E^{(\Delta 6)}$ only when the strain exceeds 3.95\%. For valleys L1, L3 and $\Delta 6$, the energy dependence on strain is nearly linear; however, a small non-linear effect can also be observed.

On the other hand, Fig.3 shows the total energy $E^{(\alpha)}$ when the strain varies from 0\% to 5\% for a film thickness of ${t}$ = 3 nm. $E^{(L1)}$ rapidly decreases as the tensile strain increases, whilst $E^{(\Delta 6)}$ increases, causing the graphs to intersect and resulting in an energy inversion. Finding the value of $\epsilon_{\|}$ such that $E^{(L1)} = E^{(\Delta 6)}$ reveals that $\epsilon_{\|}$ = 0.0388 (3.88{\%}). Therefore, for a film thickness of 3 nm, when $\epsilon_{\|}$ is greater than 0.039 (3.9{\%}), $E^{(L1)} < E^{(\Delta 6)}$ is satisfied.

\begin{figure}
\includegraphics[width=1.00\linewidth]{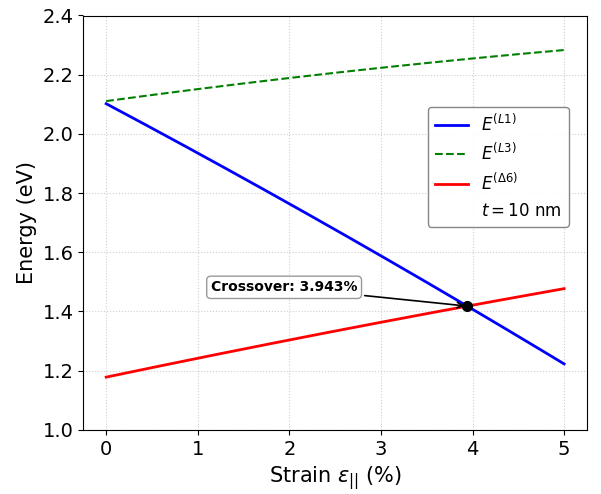}
\caption{\label{fig:epsart} The energy in each valley $E^{(\alpha)}$ when the biaxial tensile strain $\varepsilon_{\|}$ is varied from 0 to 5 \%, and the film thickness ${t}$ of the Si(111) layer is 10 nm.}
\end{figure}
\begin{figure}
\includegraphics[width=1.00\linewidth]{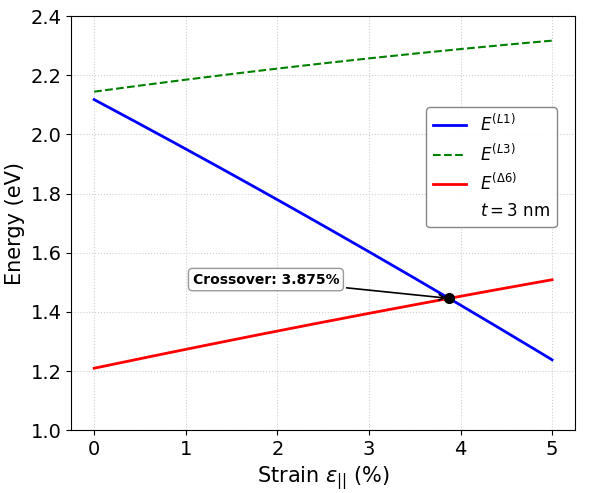}
\caption{\label{fig:epsart} The energy in each valley $E^{(\alpha)}$ when the biaxial tensile strain $\varepsilon_{\|}$ is varied from 0 to 5 \%, and the film thickness ${t}$ of the Si(111) layer is 3 nm.}
\end{figure}

As can be seen in Fig.2 and Fig.3, for a film thickness of ${t}$ = 3 nm, due to quantum effects, the total energy $E^{(L1)}$ intersects with $E^{(\Delta 6)}$ at a lower tensile stress than for ${t}$ = 10 nm. Thus, the strain $\epsilon _{\|}$ at which $E^{(L1)} = E^{(\Delta 6)}$ occurs varies with the thickness of the film ${t}$; as ${t}$ decreases, the reversal of $ E^{(L1)}$ and $E^{(\Delta 6)}$ tends to occur at smaller $\epsilon _{\|}$. Figure 4 is a graph summarizing the critical strain values over which the energy of L1 becomes lower than that of $\Delta 6$, as the thickness of the film ${t}$ is varied from 1 to 10 nm.

In Fig.4, below the crossover boundary (the region where the strain is less than the critical strain), the lowest point of the conduction band lies in the $\Delta$ valley, whereas above the crossover boundary (the region where the strain is greater than the critical strain), the lowest point of the conduction band lies in the L valley. As the thickness of the film ${t}$ approaches 1 nm, the crossover boundary increases slightly; this is thought to be because the value of $E_{q}^{(\Delta 6)}$ approaches the height of the potential barrier $V_{0}$, the rate of increase decreases, and the difference between the values of $E_{q}^{(L1)}$ and $E_{q}^{(\Delta 6)}$ narrows.

We investigate the concentration of $\text{Ge}$ ${x}$ in $\text{Si}_{1-x}\text{Ge}_{x}$ that corresponds to the critical strain in the structure $\text{Si}_{1-x}\text{Ge}_{x}$ / Si(111) / $\text{Si}_{1-x}\text{Ge}_{x}$ shown in Fig.4. 
In this structure, according to the Vegard rule, $a({x})$, the lattice constant of $\text{Si}_{1-x}\text{Ge}_{x}$ with Ge composition ${x}$ can be expressed in \AA \ as follows\cite{warlimont2018springer}:
\begin{equation}
\begin{aligned}
& a(x)=(1-x) a_{s i}+x a_{G e}+b x(1-x) \\
& a_{S i}=5.4307, \quad a_{G e}=5.6575, \quad b=-0.0273.
\end{aligned}
\end{equation}
Equation (25) and the biaxial strain in this case,
\begin{equation}
\begin{aligned}
\varepsilon_{\|}=\frac{a(x)}{a_{S i}}-1,
\end{aligned}
\end{equation}
are used to determine the Ge concentration ${x}$ in $\text{Si}_{1-x}\text{Ge}_{x}$ which realizes the critical strain such that $E^{(L1)} < E^{(\Delta 6)}$ as the thickness of the film ${t}$ is varied from 1 to 10 nm, as summarized in Fig.5.
\begin{figure}
\includegraphics[width=1.00\linewidth]{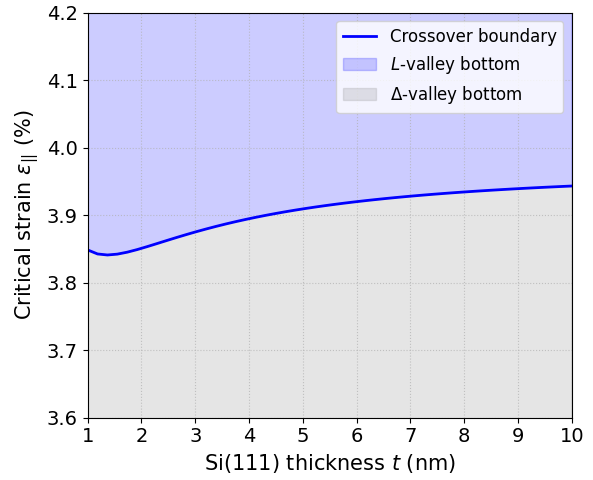}
\caption{\label{fig:epsart} Critical biaxial tensile strain $\varepsilon_{\|}$ at which the energies of L1 and $\Delta 6$ intersect, as the thickness of the film ${t}$ is varied from 1 to 10 nm.}
\end{figure}
\begin{figure}
\includegraphics[width=1.00\linewidth]{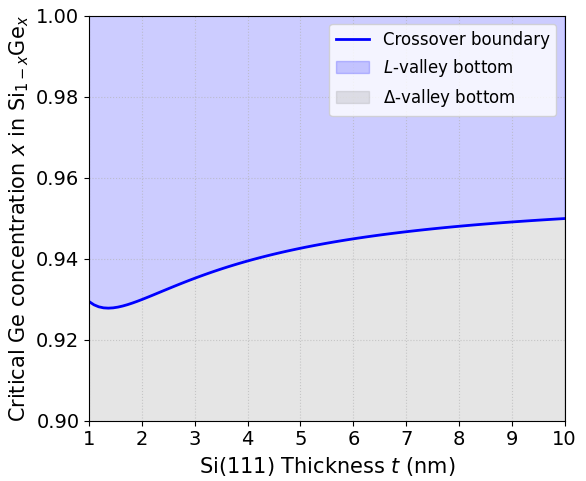}
\caption{\label{fig:epsart} The $\text{Ge}$ concentration ${x}$ in $\text{Si}_{1-x}\text{Ge}_{x}$ corresponding to the critical strain such that $E^{(L1)} < E^{(\Delta 6)}$ as the thickness of the film ${t}$ is varied from 1 to 10 nm.}
\end{figure}

For example, for a film thickness of ${t}$ = 3 nm, the critical strain at which $E^{(L1)}$ becomes smaller than $E^{(\Delta 6)}$ is 3.88\%, and the corresponding critical concentration of $\text{Ge}$ ${x}$ is 0.935. For a film thickness of ${t}$ = 4 nm, the critical $\text{Ge}$ concentration ${x}$ is 0.939; that is, for film thicknesses of 4 nm or less, when the $\text{Ge}$ concentration ${x}\geqq 0.94$, $E^{(L1)} < E^{(\Delta 6)}$. Furthermore, as can be seen from Fig.5, it is clear that, across the entire range of film thicknesses from 1 to 10 nm, electrons within the Si(111) crystal of the $\text{Ge}$ / Si(111) / $\text{Ge}$ structure where ${x}$ = 1 are confined at the L point of the band structure. 

As the Si(111) layer is sandwiched between pure Ge layers and subject to extremely large tensile strain, there are many challenges to overcome in achieving the $\text{Ge}$ / Si(111) / $\text{Ge}$ heterostructure by the deposition process; these will be summarized in the next chapter. 
\begin{figure}
\includegraphics[width=0.95\linewidth]{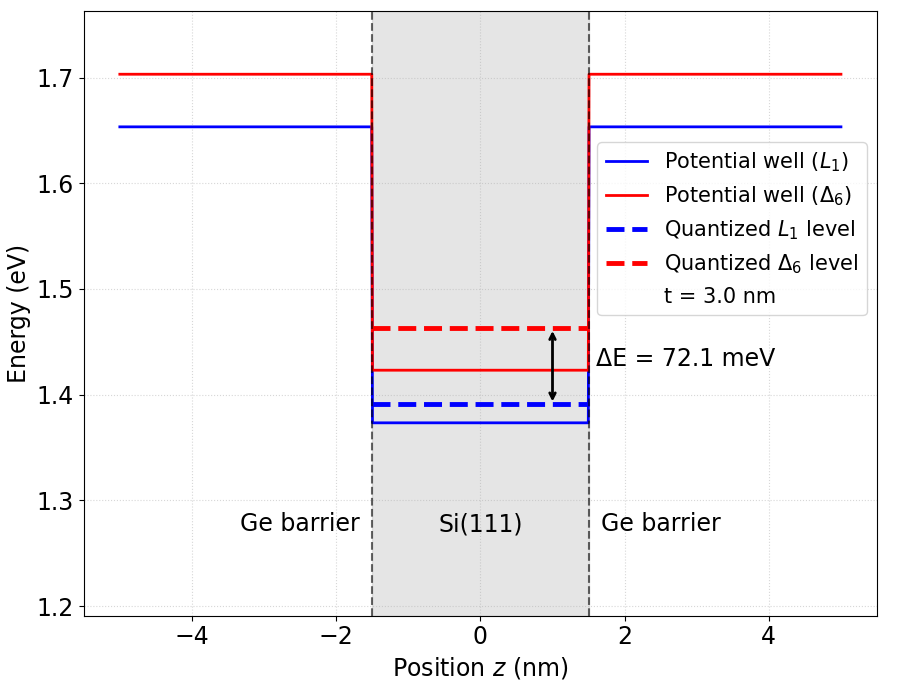}
\caption{\label{fig:epsart} The band of the Ge / Si(111) / Ge structure when the film thickness of Si(111) is 3 nm. The height of potential barrier is 0.28eV. This band structure is called Type II (staggered).}
\end{figure}

In the structure $\text{Ge}$ / Si(111) / $\text{Ge}$ for a film thickness of 3 nm, the lowest energy point is L1, and calculations show that the $\Delta 6$ point lies approximately 72.1 meV above this. Since the width of the Zeeman splitting in a qubit system is of the order of tens of $\mu$eV, the energy difference between the L1 point and the $\Delta 6$ point is sufficiently greater than this and it is considered unlikely to interfere with the two-level qubit system. 

The conduction band in this band structure is shown in Fig.6. This band structure is called Type II (staggered), although the valence band is not shown in this diagram, where electrons accumulate in the Si layer and holes accumulate in the SiGe layer, separated into different layers\cite{rieger1993electronic,yang2004si,virgilio2006type}.

\subsection{\label{sec:level2}Feasibility of Si(111) devices exceeding 3.9 percents tensile strain}

The above investigations indicate that a structure such as $\text{Si}_{1-x}\text{Ge}_{x}$ (${x}\geqq0.94$)\ / Si(111) ($\leqq 4$ nm) / $\text{Si}_{1-x}\text{Ge}_{x}$ (${x}\geqq0.94$) is necessary to achieve the lowest point of the conduction band in the L valley. The deposition process is as follows.
\begin{enumerate}
    \item Deposit a highly relaxed film of $\text{Si}_{1-x}\text{Ge}_{x}$ (${x}\geqq0.94$), extremely close to pure Ge, onto a Si(111) substrate. To achieve this, the Ge concentration is gradually increased from the surface of the Si(111) substrate, resulting in a film with (${x}\geqq0.94$) in the very top layer of SiGe. Whilst the $\text{Si}_{1-x}\text{Ge}_{x}$ film is thin, it is lattice-matched to the underlying Si(111) substrate and accumulates strain; however, as the film thickness increases, the $\text{Si}_{1-x}\text{Ge}_{x}$ film relaxes. Therefore, the total thickness of the film is deposited at a level sufficient for relaxation.\cite{bean1985strained,lee2006challenges,bolkhovityanov2007plastic,said2002design,zou2010facile,gatti2014ge}.
    \item Grow a Si(111) film with thickness ($\leqq 4$ nm) coherently on the $\text{Si}_{1-x}\text{Ge}_{x}$ (${x}\geqq0.94$) film. To achieve this, it is necessary to prevent the growth of the island and the diffusion of Ge atoms into the Si crystal\cite{tosaka2013strain}.
    \item Grow a $\text{Si}_{1-x}\text{Ge}_{x}$ (${x}\geqq0.94$) film coherently with the substrate on the Si(111) film. This film can be thin as long as it functions as a potential barrier and protective layer. 
\end{enumerate}
The above process is thought to present the following two major challenges.

\begin{enumerate}[a)]
   \item During Si(111) epitaxial growth on a $\text{Si}_{1-x}\text{Ge}_{x}$ (${x}\geqq0.94$), which is extremely close to pure Ge, the island growth (Stranski-Krastanov growth) is prone to occur. Therefore, preventing island growth is necessary to obtain a sufficiently flat and good morphology.
   \item The Si(111) crystal exhibits a tensile strain exceeding 3.9\%. Should this strain energy surpass the energy required to generate dislocations, the film relaxes, ceasing to be a strained Si film\cite{liu2022role,gradwohl2023strain,gradwohl2025enhanced}. Therefore, during the deposition of a film with thickness ($\leqq 4$ nm), it is necessary to ensure that this strain energy does not exceed the energy required to induce misfit dislocation formation.
\end{enumerate}
A detailed discussion of a) is beyond the scope of this paper; however, progress in deposition technology, supported by developments in the semiconductor industry, is expected to overcome this challenge. 
\begin{figure}
\includegraphics[width=1.00\linewidth]{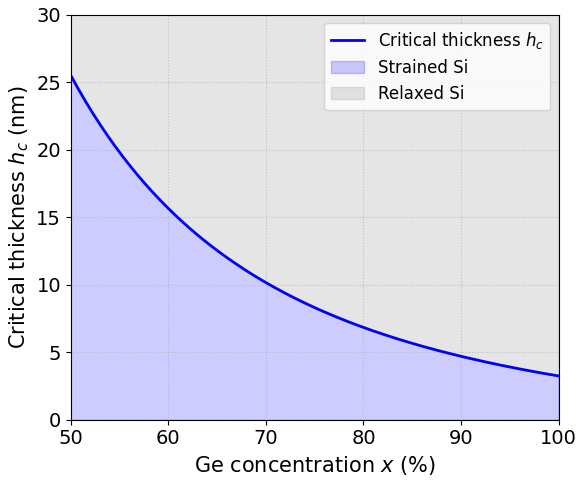}
\caption{\label{fig:epsart} The critical film thickness $h_c$ of Si(111) layer in the structure $\text{Si}_{1-x}\text{Ge}_{x}$ \ / Si(111)  / $\text{Si}_{1-x}\text{Ge}_{x}$ as a function of the Ge concentration $\text{x}$, $0.50\leqq \text{x} \leqq1.00$.}
\end{figure}
In systems with significant lattice mismatch (such as those with a high Ge concentration), island growth occurs because surface atoms attempt to release strain energy by rearranging themselves into islands before dislocations can form in the flat film. Consequently, deposition techniques operating at low temperatures, where deposited Si atoms are less mobile, are essential. Setting the growth temperature at 300°C to 400°C is believed to be effective; experimental data also report flat Si layer growth of approximately 10-12 nm within this temperature \cite{hollmann2020large,liu2022viewing,liu2023growth}. We expect progress in MBE and CVD techniques operating at this growth temperature.

Regarding issue b), the boundary thickness where strain relaxation begins is termed the critical thickness $h_{c}$, and two theoretical models are employed. The first is a model based on mechanical equilibrium, where ‘existing dislocations are stretched by strain to form misfit dislocations’; this is known as the Matthews-Blakeslee (M-B) model\cite{matthews1974defects}. The second is a model based on energy equilibrium, where ‘the energy required to generate a dislocation is equal to the strain energy’; this is known as the People-Bean model (P-B)\cite{people1985calculation}.

For the growth of Si films on $\text{Si}_{1-x}\text{Ge}_{x}$, it has been found that the latter (P-B) model agrees well with the experimental data when ${x}$ is large (high Ge concentration). Here, the (P-B) model is used for the analysis. In the P-B model, the critical film thickness $h_c$ is expressed by the following equation,
\begin{equation}
h_c=\frac{b}{32 \pi f^2} \frac{1-\nu}{1+\nu} \ln \left(\frac{h_c}{b}\right).
\end{equation}
Here, ${f}$ is the lattice mismatch ratio (Misfit) of the Si lattice constant relative to the SiGe substrate. If the concentration of Ge is denoted by ${x}$, $f \approx 0.0418  x$. ${b}$ is the magnitude of the Burgers vector, which is approximately 0.384 nm for Si. Furthermore, $\nu$ is the Poisson's ratio. The effective Poisson's ratio in the [111] direction is expressed using the elastic rigidity constants $C_{11}, C_{12}$ and $C_{44}$ as follows:
\begin{equation}
\begin{aligned}
& \nu_{111}=\frac{ R_{111}}{2+ R_{111}}, \\
& R_{111}=\frac{2\left(C_{11}+2 C_{12}-2 C_{44}\right)}{C_{11}+2 C_{12}+4 C_{44}} \approx 0.439.
\end{aligned}
\end{equation}
For Si, the elastic constants are $C_{11}=165.7 \mathrm{GPa},\  C_{12}=63.9 \mathrm{GPa},\  C_{44}=79.6 \mathrm{GPa}$\cite{ungersboeck2007effect,LandoltBornstein2001:sm_lbs_978-3-540-31355-7_212}.
Substituting these parameters, the dependence of the critical film thickness $h_c$ on the Ge concentration is calculated and shown in Fig.7.

As can be seen in Fig.7, the critical film thickness $h_c$ for the epitaxial growth of Si(111) on $\text{Si}_{1-x}\text{Ge}_{x}$ (${x}\geqq0.94$) exceeds 3 nm, and the structure $\text{Si}_{1-x}\text{Ge}_{x}$ (${x}\geqq0.94$) / strained Si(111) ($\leqq$ 3 nm) / $\text{Si}_{1-x}\text{Ge}_{x}$ (${x}\geqq0.94$) is feasible.

\section{\label{sec:level1}DISCUSSIONS AND CONCLUSIONS}

In conventional spin qubits using Si(001) crystals, electrons are confined in the $\Delta$ valley. There is a double degeneracy in its ground state, and a challenge has been that when this degeneracy unstablely lifts, it causes problems for the two-level system as a qubit.

We therefore propose to use a $\text{Si}_{1-x}\text{Ge}_{x} / \text{Si(111)} /  \text{Si}_{1-x}\text{Ge}_{x}$ structure. By utilizing the biaxial tensile strain within the (111) plane, we shift the minimum-energy state of the conduction band from the doubly degenerate valley ground state $\Delta$ to the undegenerate valley ground state L. Thus, we are led to investigate a qubit device with an electron confined in the L valley\cite{tokunaga2025111}.

It is found that shifting the minimum-energy state of the conduction band to the ground state of the L valley is achievable by maintaining the film thickness of Si(111) at 4 nm or less to prevent strain relaxation, whilst ensuring that the strain magnitude exceeds approximately 3.9\%.

In the calculations leading to the above results, standard values are used for the deformation potentials in both the linear and non-linear terms. Figure 8 shows the results obtained when a 10\% variation is introduced into the first-order deformation potentials, $\Xi_{d}$ and $\Xi_{u}$. Similarly, Fig.9 shows the results when the coefficients $D_{ijkl}$ of the second-order nonlinear term are varied, as shown in the following equations
\begin{equation}
\begin{aligned}
&\Delta E_2^{L 1}=(-15.0 \sim -30.0) \varepsilon_{\|}^2 \ \text{eV},\\
&\Delta E_2^{L 3}=(-10.0 \sim -20.0) \varepsilon_{\|}^2 \ \text{eV},\\
&\Delta E_2^{\Delta 6}=(-5.0 \sim -15.0) \varepsilon_{\|}^2 \ \text{eV}.
\end{aligned}
\end{equation}
Finally, Fig.10 shows the results obtained by taking into account variations in both the coefficients of the first-order linear terms and the coefficients of the second-order nonlinear terms. As can be seen from Fig.8 and Fig.9, the influence of the variation in the first-order deformation potential is greater than that of the variation in the second-order nonlinear term. However, when the variations in the first-order linear term and the second-order nonlinear term do not act simultaneously, it can be seen that, throughout the entire range of film thicknesses for the sandwiched Si(111) layer from 1 to 10 nm, in the structure $\text{Ge}$ / Si(111) / $\text{Ge}$, the lowest point of the conduction band energy can be shifted from the $\Delta$ valley to the L valley.

Furthermore, as can be seen from Fig.10, if the thickness of the sandwiched Si(111) layer is 4 nm or less, it is found that even when taking into account the variations in both the first-order linear term and the second-order nonlinear term of the deformation potential, adopting the structure $\text{Ge}$ / Si(111) / $\text{Ge}$ allows the lowest point of the conduction band energy to be shifted to the ground state of the L valley with a margin of safety.

The conclusions of this study are summarized below.
\begin{enumerate}
  \item When changes in the energies of the conduction band’s $\Delta$ valley and L valley due to biaxial tensile strain are calculated using standard values for both the coefficients of the linear and non-linear terms of the deformation potential, in the $\text{Si}_{1-x}\text{Ge}_{x}$ / Si(111) / $\text{Si}_{1-x}\text{Ge}_{x}$ structure, by setting the thickness of the Si(111) layer to 4 nm or less to prevent strain relaxation, whilst maintaining a strain magnitude of approximately 3.9\% or more—i.e., by setting the Ge concentration \(x > 0.94\)—it is possible to shift the lowest point of the conduction band energy to the ground state of the L-valley.

  \item When changes in the energies of the conduction band’s $\Delta$ valley and L valley due to biaxial tensile strain are calculated taking into account the maximum possible variation in the coefficients of the linear and non-linear terms of the deformation potential, in the structure $\text{Ge}$ / Si(111) / $\text{Ge}$, by setting the thickness of the Si(111) layer to 4 nm or less to prevent strain relaxation, it is possible to shift the lowest point of the conduction band energy from the $\Delta$ valley to the L valley. 

  \item To achieve a strain magnitude of approximately 3.9$\% $ or greater, the structure $\text{Si}_{1-x}\text{Ge}_{x}$ / Si(111) / $\text{Si}_{1-x}\text{Ge}_{x}$ requires a high concentration of Ge ${x}\geqq0.94$. Consequently, technological development is necessary to prevent island growth and deposit sufficiently flat films during the epitaxial growth of the Si(111) layer. For the direction of this technological development, we expect progress using MBE and CVD techniques with growth temperatures of 300°C to 400°C.

  \item Additionally, it is necessary to eliminate high-temperature thermal processing steps to prevent the diffusion of Ge atoms into the interposed Si(111) layer after forming the structure $\text{Si}_{1-x}\text{Ge}_{x}$ (${x}\geqq0.94$) / Si(111) ($\leqq$ 4 nm) / $\text{Si}_{1-x}\text{Ge}_{x}$ (${x}\geqq0.94$). When thermal processing is required, processes with low thermal loading on the substrate, such as Rapid Thermal Annealing (RTA), are suitable.

  \item Consequently, when qubit devices are integrated with conventional CMOS devices, it is anticipated that either the CMOS devices must be formed beforehand with the qubit devices fabricated subsequently, or the qubit devices and CMOS devices must be manufactured completely separately and then integrated later. In such cases, it is desirable that wafers with the structure $\text{Si}_{1-x}\text{Ge}_{x}$ (${x}\geqq0.94$) / Si(111) ($\leqq$ 4 nm) / $\text{Si}_{1-x}\text{Ge}_{x}$ (${x}\geqq0.94$) are prepared for qubit devices.

  \item The effective mass of electrons confined in the L valley of Si(111) in the transverse $( x, y )$ direction is very low, approximately $0.12m_{0}$. Furthermore, as the ground state L1 of the L valley is non-degenerate, although this is beyond the scope of this paper, it is thought that the effect of valley scattering on the lateral movement of electrons is small. Consequently, in Field Effect Transistors (FET) and similar technologies that utilize lateral $( x, y )$ movement of electrons confined in the L valley of Si(111), the electron mobility is very high and such devices are expected to be promising as extremely high-speed device technologies. 
\end{enumerate}
\begin{figure}
\includegraphics[width=1.00\linewidth]{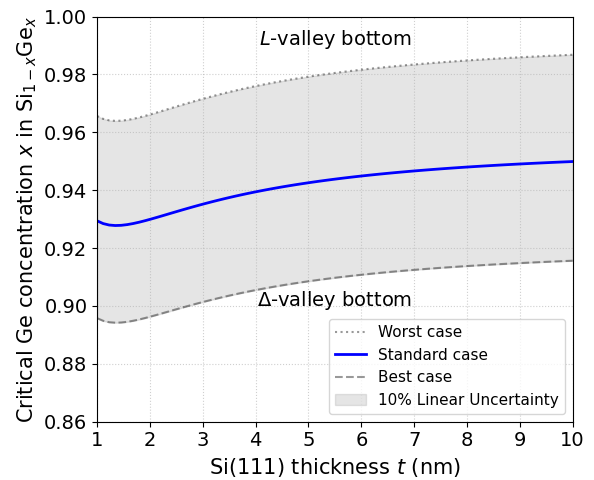}
\caption{\label{fig:epsart} The $\text{Ge}$ concentration ${x}$ in $\text{Si}_{1-x}\text{Ge}_{x}$ corresponding to the critical strain such that $E^{(L1)} < E^{(\Delta 6)}$ when a 10\%  variation is introduced into the first-order deformation potentials, $\Xi_{d}$ and $\Xi_{u}$.}
\end{figure}

\begin{figure}
\includegraphics[width=1.00\linewidth]{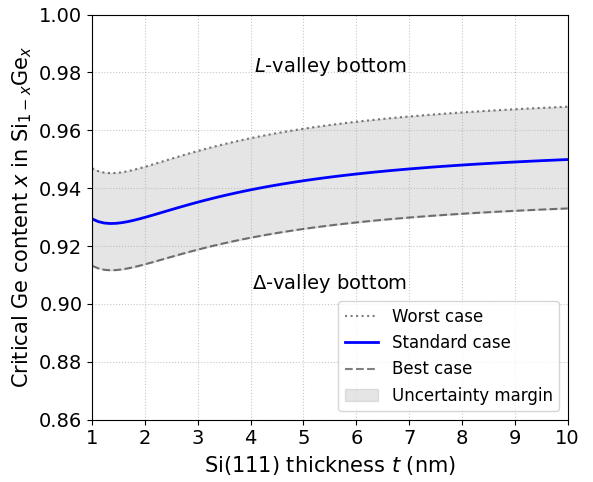}
\caption{\label{fig:epsart}The $\text{Ge}$ concentration ${x}$ in $\text{Si}_{1-x}\text{Ge}_{x}$ corresponding to the critical strain such that $E^{(L1)} < E^{(\Delta 6)}$ when the coefficients $D_{ijkl}$ of the second-order nonlinear term are varied.}
\end{figure}

\begin{figure}
\includegraphics[width=1.00\linewidth]{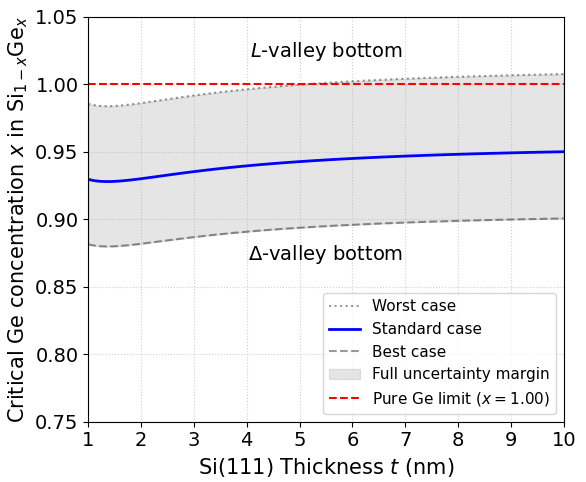}
\caption{\label{fig:epsart} The $\text{Ge}$ concentration ${x}$ in $\text{Si}_{1-x}\text{Ge}_{x}$ corresponding to the critical strain such that $E^{(L1)} < E^{(\Delta 6)}$ when the variation in both of the first-order deformation potentials and the second-order nonlinear term is introduced.}.
\end{figure}
\begin{acknowledgments}
The authors thank the members of our laboratory for their helpful discussions and support throughout this work.
\end{acknowledgments}
\section*{AUTHOR DECLARATIONS}
\subsection*{Conflict of Interest}
Takafumi Tokunaga and Hiromichi Nakazato have patents pending related to the methods described in this manuscript.

\subsection*{Author contributions}
\textbf{Takafumi Tokunaga}: Conceptualization (lead); Methodology (lead); Data curation (lead); Formal analysis (equal); Resources (equal); Visualization (equal); Writing – original draft (lead); Writing–review and editing (equal).
\textbf{Hiromichi Nakazato}: Funding acquisition (lead); Conceptualization (equal); Methodology (equal); Data curation (equal); Formal analysis (equal); Resources (lead); Visualization (equal); Writing–review and editing (lead); Project administration (lead); Supervision (lead).

\subsection*{Data Availability}
Data supporting the findings of this study are available from the corresponding author on a reasonable request.
\section*{REFERENCES}
\nocite{*}
\bibliographystyle{ieeetr}
\bibliography{references}

@PREAMBLE{
 "\providecommand{\noopsort}[1]{}" 
 # "\providecommand{\singleletter}[1]{#1}%" 
}

@article{maurand2016cmos,
  title={{A CMOS silicon spin qubit}},
  author={Maurand, R and Jehl, X and Kotekar-Patil, D and Corna, Andrea and Bohuslavskyi, Heorhii and Lavi{\'e}ville, R and Hutin, L and Barraud, S and Vinet, M and Sanquer, M and others},
  journal={Nature Communications},
  volume={7},
  number={1},
  pages={13575},
  year={2016},
  publisher={Nature Publishing Group UK London}
}

@article{camenzind2022hole,
  title={{A hole spin qubit in a fin field-effect transistor above 4 Kelvin}},
  author={Camenzind, Leon C and Geyer, Simon and Fuhrer, Andreas and Warburton, Richard J and Zumb{\"u}hl, Dominik M and Kuhlmann, Andreas V},
  journal={Nature Electronics},
  volume={5},
  number={3},
  pages={178--183},
  year={2022},
  publisher={Nature Publishing Group UK London}
}

@article{klemt2023electrical,
  title={{Electrical manipulation of a single electron spin in CMOS using a micromagnet and spin-valley coupling}},
  author={Klemt, Bernhard and Elhomsy, Victor and Nurizzo, Martin and Hamonic, Pierre and Martinez, Biel and Cardoso Paz, Bruna and Spence, Cameron and Dartiailh, Matthieu C and Jadot, Baptiste and Chanrion, Emmanuel and others},
  journal={npj Quantum Information},
  volume={9},
  number={1},
  pages={107},
  year={2023},
  publisher={Nature Publishing Group UK London}
}

@article{geyer2024anisotropic,
  title={{Anisotropic exchange interaction of two hole-spin qubits}},
  author={Geyer, Simon and Het{\'e}nyi, Bence and Bosco, Stefano and Camenzind, Leon C and Eggli, Rafael S and Fuhrer, Andreas and Loss, Daniel and Warburton, Richard J and Zumb{\"u}hl, Dominik M and Kuhlmann, Andreas V},
  journal={Nature Physics},
  volume={20},
  number={7},
  pages={1152--1157},
  year={2024},
  publisher={Nature Publishing Group UK London}
}

@article{steinacker2025industry,
  title={{Industry-compatible silicon spin-qubit unit cells exceeding 99\% fidelity}},
  author={Steinacker, Paul and Dumoulin Stuyck, Nard and Lim, Wee Han and Tanttu, Tuomo and Feng, MengKe and Serrano, Santiago and Nickl, Andreas and Candido, Marco and Cifuentes, Jesus D and Vahapoglu, Ensar and others},
  journal={Nature},
  volume={646},
  number={8083},
  pages={81--87},
  year={2025},
  publisher={Nature Publishing Group UK London}
}

@article{friesen2007valley,
  title={{Valley splitting theory of SiGe/ Si/ SiGe quantum wells}},
  author={Friesen, Mark and Chutia, Sucismita and Tahan, Charles and Coppersmith, SN},
  journal={Physical Review B—Condensed Matter and Materials Physics},
  volume={75},
  number={11},
  pages={115318},
  year={2007},
  publisher={APS}
}

@article{saraiva2009physical,
  title={{Physical mechanisms of interface-mediated intervalley coupling in Si}},
  author={Saraiva, AL and Calder{\'o}n, MJ and Hu, Xuedong and Das Sarma, Sankar and Koiller, Belita},
  journal={Physical Review B—Condensed Matter and Materials Physics},
  volume={80},
  number={8},
  pages={081305},
  year={2009},
  publisher={APS}
}

@article{saraiva2011intervalley,
  title={{Intervalley coupling for interface-bound electrons in silicon: An effective mass study}},
  author={Saraiva, AL and Calder{\'o}n, MJ and Capaz, Rodrigo B and Hu, Xuedong and Das Sarma, S and Koiller, Belita},
  journal={Physical Review B—Condensed Matter and Materials Physics},
  volume={84},
  number={15},
  pages={155320},
  year={2011},
  publisher={APS}
}

@article{paquelet2022atomic,
  title={{Atomic fluctuations lifting the energy degeneracy in Si/SiGe quantum dots}},
  author={Paquelet Wuetz, Brian and Losert, Merritt P and Koelling, Sebastian and Stehouwer, Lucas EA and Zwerver, Anne-Marije J and Philips, Stephan GJ and M{\k{a}}dzik, Mateusz T and Xue, Xiao and Zheng, Guoji and Lodari, Mario and others},
  journal={Nature Communications},
  volume={13},
  number={1},
  pages={7730},
  year={2022},
  publisher={Nature Publishing Group UK London}
}

@article{losert2023practical,
  title={{Practical strategies for enhancing the valley splitting in Si/SiGe quantum wells}},
  author={Losert, Merritt P and Eriksson, MA and Joynt, Robert and Rahman, Rajib and Scappucci, Giordano and Coppersmith, Susan N and Friesen, Mark},
  journal={Physical Review B},
  volume={108},
  number={12},
  pages={125405},
  year={2023},
  publisher={APS}
}

@article{lima2023interface,
  title={{Interface and electromagnetic effects in the valley splitting of Si quantum dots}},
  author={Lima, Jonas RF and Burkard, Guido},
  journal={Materials for Quantum Technology},
  volume={3},
  number={2},
  pages={025004},
  year={2023},
  publisher={IOP Publishing}
}

@article{borselli2011measurement,
  title={{Measurement of valley splitting in high-symmetry Si/SiGe quantum dots}},
  author={Borselli, Matthew G and Ross, Richard S and Kiselev, Andrey A and Croke, Edward T and Holabird, Kevin S and Deelman, Peter W and Warren, Leslie D and Alvarado-Rodriguez, Ivan and Milosavljevic, Ivan and Ku, Fiona C and others},
  journal={Applied Physics Letters},
  volume={98},
  number={12},
  year={2011},
  publisher={AIP Publishing}
}

@article{neyens2018critical,
  title={{The critical role of substrate disorder in valley splitting in Si quantum wells}},
  author={Neyens, Samuel F and Foote, Ryan H and Thorgrimsson, Brandur and Knapp, TJ and McJunkin, Thomas and Vandersypen, LMK and Amin, Payam and Thomas, Nicole K and Clarke, James S and Savage, DE and others},
  journal={Applied Physics Letters},
  volume={112},
  number={24},
  year={2018},
  publisher={AIP Publishing}
}

@article{hollmann2020large,
  title={{Large, tunable valley splitting and single-spin relaxation mechanisms in a $\text{Si}/\text{Si}_{x}\text{Ge}_{1-x}$ quantum dot}},
  author={Hollmann, Arne and Struck, Tom and Langrock, Veit and Schmidbauer, Andreas and Schauer, Floyd and Leonhardt, Tim and Sawano, Kentarou and Riemann, Helge and Abrosimov, Nikolay V and Bougeard, Dominique and others},
  journal={Physical Review Applied},
  volume={13},
  number={3},
  pages={034068},
  year={2020},
  publisher={APS}
}

@article{degli2024low,
  title={{Low disorder and high valley splitting in silicon}},
  author={Degli Esposti, Davide and Stehouwer, Lucas EA and G{\"u}l, {\"O}nder and Samkharadze, Nodar and D{\'e}prez, Corentin and Meyer, Marcel and Meijer, Ilja N and Tryputen, Larysa and Karwal, Saurabh and Botifoll, Marc and others},
  journal={npj Quantum Information},
  volume={10},
  number={1},
  pages={32},
  year={2024},
  publisher={Nature Publishing Group UK London}
}

@article{yang2013spin,
  title={{Spin-valley lifetimes in a silicon quantum dot with tunable valley splitting}},
  author={Yang, CH and Rossi, A and Ruskov, R and Lai, NS and Mohiyaddin, FA and Lee, S and Tahan, C and Klimeck, Gerhard and Morello, A and Dzurak, AS},
  journal={Nature Communications},
  volume={4},
  number={1},
  pages={2069},
  year={2013},
  publisher={Nature Publishing Group UK London}
}

@article{borjans2019single,
  title={{Single-spin relaxation in a synthetic spin-orbit field}},
  author={Borjans, F and Zajac, DM and Hazard, TM and Petta, JR},
  journal={Physical Review Applied},
  volume={11},
  number={4},
  pages={044063},
  year={2019},
  publisher={APS}
}

@article{losert2024strategies,
  title={{Strategies for enhancing spin-shuttling fidelities in Si/Si Ge quantum wells with random-alloy disorder}},
  author={Losert, Merritt P and Oberl{\"a}nder, Max and Teske, Julian D and Volmer, Mats and Schreiber, Lars R and Bluhm, Hendrik and Coppersmith, SN and Friesen, Mark},
  journal={PRX Quantum},
  volume={5},
  number={4},
  pages={040322},
  year={2024},
  publisher={APS}
}

@article{david2024long,
  title={{Long distance spin shuttling enabled by few-parameter velocity optimization}},
  author={David, Alessandro and Pazhedath, Akshay Menon and Schreiber, Lars R and Calarco, Tommaso and Bluhm, Hendrik and Motzoi, Felix},
  journal={arXiv preprint arXiv:2409.07600},
  year={2024}
}

@article{feng2022enhanced,
  title={{Enhanced valley splitting in Si layers with oscillatory Ge concentration}},
  author={Feng, Yi and Joynt, Robert},
  journal={Physical Review B},
  volume={106},
  number={8},
  pages={085304},
  year={2022},
  publisher={APS}
}

@article{woods2024coupling,
  title={{Coupling conduction-band valleys in SiGe heterostructures via shear strain and Ge concentration oscillations}},
  author={Woods, Benjamin D and Soomro, Hudaiba and Joseph, ES and Frink, Collin CD and Joynt, Robert and Eriksson, MA and Friesen, Mark},
  journal={npj Quantum Information},
  volume={10},
  number={1},
  pages={54},
  year={2024},
  publisher={Nature Publishing Group UK London}
}

@article{mcjunkin2022sige,
  title={{SiGe quantum wells with oscillating Ge concentrations for quantum dot qubits}},
  author={McJunkin, Thomas and Harpt, Benjamin and Feng, Yi and Losert, Merritt P and Rahman, Rajib and Dodson, JP and Wolfe, MA and Savage, DE and Lagally, MG and Coppersmith, SN and others},
  journal={Nature Communications},
  volume={13},
  number={1},
  pages={7777},
  year={2022},
  publisher={Nature Publishing Group UK London}
}

@article{tokunaga2025111,
  title={{(111) Si spin qubits constructed on L point of band structure}},
  author={Tokunaga, Takafumi and Nakazato, Hiromichi},
  journal={arXiv preprint arXiv:2501.13546},
  year={2025}
}

@article{bardeen1950deformation,
  title={{Deformation potentials and mobilities in non-polar crystals}},
  author={Bardeen, John and Shockley, WJPR},
  journal={Physical Review},
  volume={80},
  number={1},
  pages={72},
  year={1950},
  publisher={APS}
}

@article{herring1956transport,
  title={{Transport and deformation-potential theory for many-valley semiconductors with anisotropic scattering}},
  author={Herring, Conyers and Vogt, Erich},
  journal={Physical Review},
  volume={101},
  number={3},
  pages={944},
  year={1956},
  publisher={APS}
}

@article{van1985theoretical,
  title={{Theoretical study of Si/Ge interfaces}},
  author={Van de Walle, Chris G and Martin, Richard M},
  journal={Journal of Vacuum Science \& Technology B: Microelectronics Processing and Phenomena},
  volume={3},
  number={4},
  pages={1256--1259},
  year={1985},
  publisher={American Vacuum Society}
}

@article{van1986theoretical,
  title={{Theoretical calculations of heterojunction discontinuities in the Si/Ge system}},
  author={Van de Walle, Chris G and Martin, Richard M},
  journal={Physical Review B},
  volume={34},
  number={8},
  pages={5621},
  year={1986},
  publisher={APS}
}

@article{rieger1993electronic,
  title={{Electronic-band parameters in strained $\text{Si}_{1-x}\text{Ge}_{x}$ alloys on $\text{Si}_{1-y}\text{Ge}_{y}$ substrates}},
  author={Rieger, Martin M and Vogl, Peter},
  journal={Physical Review B},
  volume={48},
  number={19},
  pages={14276},
  year={1993},
  publisher={APS}
}

@article{hinckley1990influence,
  title={{Influence of substrate composition and crystallographic orientation on the band structure of pseudomorphic Si-Ge alloy films}},
  author={Hinckley, JM and Singh, J},
  journal={Physical Review B},
  volume={42},
  number={6},
  pages={3546},
  year={1990},
  publisher={APS}
}

@article{fischetti1996band,
  title={{Band structure, deformation potentials, and carrier mobility in strained Si, Ge, and SiGe alloys}},
  author={Fischetti, Max V and Laux, Steven E},
  journal={Journal of Applied Physics},
  volume={80},
  number={4},
  pages={2234--2252},
  year={1996},
  publisher={American Institute of Physics}
}

@article{li1991shear,
  title={{Shear-deformation-potential constant of the conduction-band minima of Si: Experimental determination by the deep-level capacitance transient method}},
  author={Li, Ming-Fu and Zhao, Xue-Shu and Gu, Zong-Quan and Chen, Jian-Xin and Li, Yan-Jin and Wang, Jian-Qing},
  journal={Physical Review B},
  volume={43},
  number={17},
  pages={14040},
  year={1991},
  publisher={APS}
}

@article{ungersboeck2007effect,
  title={{The effect of general strain on the band structure and electron mobility of silicon}},
  author={Ungersboeck, Enzo and Dhar, Siddhartha and Karlowatz, Gerhard and Sverdlov, Viktor and Kosina, Hans and Selberherr, Siegfried},
  journal={IEEE Transactions on Electron Devices},
  volume={54},
  number={9},
  pages={2183--2190},
  year={2007},
  publisher={IEEE}
}

@article{van1989band,
  title={{Band lineups and deformation potentials in the model-solid theory}},
  author={Van de Walle, Chris G},
  journal={Physical Review B},
  volume={39},
  number={3},
  pages={1871},
  year={1989},
  publisher={APS}
}

@article{friedel1989local,
  title={{Local empirical pseudopotential approach to the optical properties of Si/Ge superlattices}},
  author={Friedel, P and Hybertsen, MS and Schl{\"u}ter, M},
  journal={Physical Review B},
  volume={39},
  number={11},
  pages={7974},
  year={1989},
  publisher={APS}
}

@article{fischetti1988monte,
  title={{Monte Carlo analysis of electron transport in small semiconductor devices including band-structure and space-charge effects}},
  author={Fischetti, Massimo V and Laux, Steven E},
  journal={Physical Review B},
  volume={38},
  number={14},
  pages={9721},
  year={1988},
  publisher={APS}
}

@article{yang2004si,
  title={{Si/SiGe heterostructure parameters for device simulations}},
  author={Yang, Lianfeng and Watling, Jeremy R and Wilkins, Richard CW and Borici, Mirela and Barker, John R and Asenov, Asen and Roy, Scott},
  journal={Semiconductor Science and Technology},
  volume={19},
  number={10},
  pages={1174--1182},
  year={2004}
}

@article{li2021deformation,
  title={{Deformation potential extraction and computationally efficient mobility calculations in silicon from first principles}},
  author={Li, Zhen and Graziosi, Patrizio and Neophytou, Neophytos},
  journal={Physical Review B},
  volume={104},
  number={19},
  pages={195201},
  year={2021},
  publisher={APS}
}

@article{yang2024uncovering,
  title={{Uncovering the important role of transverse acoustic phonons in the carrier-phonon scattering in silicon}},
  author={Yang, Qiao-Lin and Li, Wu and Wang, Zhi and Ning, Fan-long and Luo, Jun-Wei},
  journal={Physical Review B},
  volume={109},
  number={12},
  pages={125203},
  year={2024},
  publisher={APS}
}

@article{williams2025improved,
  title={{Improved Calculation of Acoustic Deformation Potentials from First Principles}},
  author={Williams, Patrick and Dyson, Angela},
  journal={arXiv preprint arXiv:2502.08538},
  year={2025}
}

@book{yoder1994first,
  title={{First principles Monte Carlo simulation of charge transport in semiconductors}},
  author={Yoder, Paul Douglas},
  year={1994},
  publisher={University of Illinois at Urbana-Champaign}
}

@article{tserbak1993unified,
  title={{Unified approach to the electronic structure of strained Si/Ge superlattices}},
  author={Tserbak, C and Polatoglou, HM and Theodorou, G},
  journal={Physical Review B},
  volume={47},
  number={12},
  pages={7104},
  year={1993},
  publisher={APS}
}

@article{schmid1990calculated,
  title={{Calculated deformation potentials in Si, Ge, and GeSi}},
  author={Schmid, U and Christensen, NE and Cardona, M},
  journal={Solid State Communications},
  volume={75},
  number={1},
  pages={39--43},
  year={1990},
  publisher={Elsevier}
}

@article{laude1971effects,
  title={{Effects of uniaxial stress on the indirect exciton spectrum of silicon}},
  author={Laude, Lucien D and Pollak, Fred H and Cardona, Manuel},
  journal={Physical Review B},
  volume={3},
  number={8},
  pages={2623},
  year={1971},
  publisher={APS}
}

@article{chelikowsky1976nonlocal,
  title={{Nonlocal pseudopotential calculations for the electronic structure of eleven diamond and zinc-blende semiconductors}},
  author={Chelikowsky, James R and Cohen, Marvin L},
  journal={Physical Review B},
  volume={14},
  number={2},
  pages={556},
  year={1976},
  publisher={APS}
}

@article{rideau2006strained,
  title={{Strained Si, Ge, and $\text{Si}_{1-x}\text{Ge}_{x}$ alloys modeled with a first-principles-optimized full-zone k$\cdot$p method}},
  author={Rideau, D and Feraille, M and Ciampolini, L and Minondo, M and Tavernier, C and Jaouen, H and Ghetti, A},
  journal={Physical Review B—Condensed Matter and Materials Physics},
  volume={74},
  number={19},
  pages={195208},
  year={2006},
  publisher={APS}
}

@article{chen2011conduction,
  title={{Conduction band structure and electron mobility in uniaxially strained Si via externally applied strain in nanomembranes}},
  author={Chen, Feng and Euaruksakul, Chanan and Liu, Zheng and Himpsel, FJ and Liu, Feng and Lagally, Max G},
  journal={Journal of Physics D: Applied Physics},
  volume={44},
  number={32},
  pages={325107},
  year={2011}
}

@article{canali1975electron,
  title={{Electron drift velocity in silicon}},
  author={Canali, C and Jacoboni, C and Nava, F and Ottaviani, G and Alberigi-Quaranta, A},
  journal={Physical Review B},
  volume={12},
  number={6},
  pages={2265},
  year={1975},
  publisher={APS}
}

@book{tomizawa1993numerical,
  title={{Numerical simulation of submicron semiconductor devices}},
  author={Tomizawa, Kazutaka},
  year={1993},
  publisher={Artech House}
}

@article{nielsen1985stresses,
  title={{Stresses in semiconductors: Ab initio calculations on Si, Ge, and GaAs}},
  author={Nielsen, OH and Martin, Richard M},
  journal={Physical Review B},
  volume={32},
  number={6},
  pages={3792},
  year={1985},
  publisher={APS}
}

@article{kleinman1962deformation,
  title={{Deformation potentials in silicon. I. Uniaxial strain}},
  author={Kleinman, Leonard},
  journal={Physical Review},
  volume={128},
  number={6},
  pages={2614},
  year={1962},
  publisher={APS}
}

@article{kleinman1963deformation,
  title={{Deformation potentials in silicon. II. Hydrostatic strain and the electron-phonon interaction}},
  author={Kleinman, Leonard},
  journal={Physical Review},
  volume={130},
  number={6},
  pages={2283},
  year={1963},
  publisher={APS}
}

@article{goroff1963deformation,
  title={{Deformation potentials in silicon. III. Effects of a general strain on conduction and valence levels}},
  author={Goroff, Iza and Kleinman, Leonard},
  journal={Physical Review},
  volume={132},
  number={3},
  pages={1080},
  year={1963},
  publisher={APS}
}

@article{macfarlane1958fine,
  title={{Fine structure in the absorption-edge spectrum of Si}},
  author={MacFarlane, 25\_GG and McLean, TP and Quarrington, JE and Roberts, V},
  journal={Physical Review},
  volume={111},
  number={5},
  pages={1245},
  year={1958},
  publisher={APS}
}

@article{bludau1974temperature,
  title={{Temperature dependence of the band gap of silicon}},
  author={Bludau, W and Onton, A and Heinke, W},
  journal={Journal of Applied Physics},
  volume={45},
  number={4},
  pages={1846--1848},
  year={1974},
  publisher={American Institute of Physics}
}

@article{richard2004energy,
  title={{Energy-band structure of Ge, Si, and GaAs: A thirty-band k$\cdot$p method}},
  author={Richard, Soline and Aniel, Fr{\'e}d{\'e}ric and Fishman, Guy},
  journal={Physical Review B—Condensed Matter and Materials Physics},
  volume={70},
  number={23},
  pages={235204},
  year={2004},
  publisher={APS}
}

@article{chen2010quantum,
  title={{Quantum confinement, surface roughness, and the conduction band structure of ultrathin silicon membranes}},
  author={Chen, Feng and Ramayya, Edwin B and Euaruksakul, Chanan and Himpsel, Franz J and Celler, George K and Ding, Bingjun and Knezevic, Irena and Lagally, Max G},
  journal={ACS nano},
  volume={4},
  number={4},
  pages={2466--2474},
  year={2010},
  publisher={ACS Publications}
}

@article{paul2004si,
  title={{Si/SiGe heterostructures: from material and physics to devices and circuits}},
  author={Paul, Douglas J},
  journal={Semiconductor Science and Technology},
  volume={19},
  number={10},
  pages={R75--R108},
  year={2004}
}

@article{sverdlov2008two,
  title={{Two-band k$\cdot$p model for the conduction band in silicon: Impact of strain and confinement on band structure and mobility}},
  author={Sverdlov, Viktor and Karlowatz, Gerhard and Dhar, Siddhartha and Kosina, Hans and Selberherr, Siegfried},
  journal={Solid-State Electronics},
  volume={52},
  number={10},
  pages={1563--1568},
  year={2008},
  publisher={Elsevier}
}

@article{bendaniel1966space,
  title={{Space-charge effects on electron tunneling}},
  author={BenDaniel, DJ and Duke, CB},
  journal={Physical Review},
  volume={152},
  number={2},
  pages={683},
  year={1966},
  publisher={APS}
}

@book{warlimont2018springer,
  title={{Springer handbook of materials data}},
  author={Warlimont, Hans and Martienssen, Werner},
  year={2018},
  publisher={Springer}
}

@article{virgilio2006type,
  title={{Type-I alignment and direct fundamental gap in SiGe based heterostructures}},
  author={Virgilio, Michele and Grosso, Giuseppe},
  journal={Journal of Physics: Condensed Matter},
  volume={18},
  number={3},
  pages={1021--1031},
  year={2006}
}

@article{gatti2014ge,
  title={{Ge/SiGe quantum wells on Si(111): Growth, structural, and optical properties}},
  author={Gatti, E and Isa, Fabio and Chrastina, Daniel and M{\"u}ller Gubler, E and Pezzoli, Fabio and Grilli, E and Isella, Giovanni},
  journal={Journal of Applied Physics},
  volume={116},
  number={4},
  year={2014},
  publisher={AIP Publishing}
}

@article{bean1985strained,
  title={{Strained-layer epitaxy of germanium-silicon alloys}},
  author={Bean, John C},
  journal={Science},
  volume={230},
  number={4722},
  pages={127--131},
  year={1985},
  publisher={American Association for the Advancement of Science}
}

@article{bolkhovityanov2007plastic,
  title={{Plastic relaxation of GeSi/Si(001) films grown by molecular-beam epitaxy in the presence of the Sb surfactant}},
  author={Bolkhovityanov, Yu B and Deryabin, AS and Gutakovski{\u\i}, AK and Kolesnikov, AV and Sokolov, LV},
  journal={Semiconductors},
  volume={41},
  number={10},
  pages={1234--1239},
  year={2007},
  publisher={Springer}
}

@article{said2002design,
  title={{Design, fabrication, and analysis of crystalline Si-SiGe heterostructure thin-film solar cells}},
  author={Said, Khalid and Poortmans, Jozef and Caymax, Matty and Nijs, Johan F and Debarge, Luc and Christoffel, Eric and Slaoui, Abdelilah},
  journal={IEEE Transactions on Electron Devices},
  volume={46},
  number={10},
  pages={2103--2110},
  year={2002},
  publisher={IEEE}
}

@article{lee2006challenges,
  title={{Challenges in epitaxial growth of SiGe buffers on Si(111),(110), and (112)}},
  author={Lee, Minjoo L and Antoniadis, Dimitri A and Fitzgerald, Eugene A},
  journal={Thin Solid Films},
  volume={508},
  number={1-2},
  pages={136--139},
  year={2006},
  publisher={Elsevier}
}

@article{zou2010facile,
  title={{Facile chemical solution deposition of high-mobility epitaxial germanium films on silicon}},
  author={Zou, Guifu and Luo, Hongmei and Ronning, Filip and Sun, Baoquan and McCleskey, Thomas M and Burrell, Anthony K and Bauer, Eve and Jia, QX},
  journal={Angew. Chem. Int. Ed},
  volume={49},
  number={10},
  pages={1782--1785},
  year={2010}
}

@article{tosaka2013strain,
  title={{Strain induced intermixing of Ge atoms in Si epitaxial layer on Ge(111)}},
  author={Tosaka, Aki and Mochizuki, Izumi and Negishi, Ryota and Shigeta, Yukichi},
  journal={Journal of Applied Physics},
  volume={113},
  number={7},
  year={2013},
  publisher={AIP Publishing}
}

@article{liu2022role,
  title={{Role of critical thickness in SiGe/Si/SiGe heterostructure design for qubits}},
  author={Liu, Yujia and Gradwohl, Kevin-P and Lu, Chen-Hsun and Remmele, Thilo and Yamamoto, Yuji and Zoellner, Marvin H and Schroeder, Thomas and Boeck, Torsten and Amari, Houari and Richter, Carsten and others},
  journal={Journal of Applied Physics},
  volume={132},
  number={8},
  year={2022},
  publisher={AIP Publishing}
}

@article{gradwohl2023strain,
  title={{Strain relaxation of Si/SiGe heterostructures by a geometric Monte Carlo approach}},
  author={Gradwohl, Kevin-P and Lu, Chen-Hsun and Liu, Yujia and Richter, Carsten and Boeck, Torsten and Martin, Jens and Albrecht, Martin},
  journal={Physica Status Solidi (RRL)--Rapid Research Letters},
  volume={17},
  number={6},
  pages={2200398},
  year={2023},
  publisher={Wiley Online Library}
}

@article{gradwohl2025enhanced,
  title={{Enhanced nanoscale Ge concentration oscillations in Si/SiGe quantum well through controlled segregation}},
  author={Gradwohl, Kevin-P and Cvitkovich, Lukas and Lu, Chen-Hsun and Koelling, Sebastian and Oezkent, Maximilian and others},
  journal={Nano Letters},
  volume={25},
  number={11},
  pages={4204--4210},
  year={2025},
  publisher={ACS Publications}
}

@article{liu2022viewing,
  title={{Viewing SiGe Heterostructure for Qubits with Dislocation Theory}},
  author={Liu, Yujia and Gradwohl, Kevin-Peter and Lu, Chenhsun and Yamamoto, Yuji and Remmele, Thilo and Corley-Wiciak, Cedric and Teubner, Thomas and Richter, Carsten and Albrecht, Martin and Boeck, Torsten},
  journal={Electrochemical Society Transactions 242},
  volume={109},
  number={4},
  pages={189--196},
  year={2022},
  publisher={The Electrochemical Society, Inc.}
}

@article{liu2023growth,
  title={{Growth of 28Si Quantum Well Layers for Qubits by a Hybrid MBE/CVD Technique}},
  author={Liu, Yujia and Gradwohl, Kevin-Peter and Lu, Chen-HSun and Yamamoto, Yuji and Remmele, Thilo and Corley-Wiciak, Cedric and Teubner, Thomas and Richter, Carsten and Albrecht, Martin and Boeck, Torsten},
  journal={ECS Journal of Solid State Science and Technology},
  volume={12},
  number={2},
  pages={024006},
  year={2023},
  publisher={IOP Publishing}
}

@article{matthews1974defects,
  title={{Defects in epitaxial multilayers: I. Misfit dislocations}},
  author={Matthews, John W and Blakeslee, A Eugene},
  journal={Journal of Crystal Growth},
  volume={27},
  pages={118--125},
  year={1974},
  publisher={Elsevier}
}

@article{people1985calculation,
  title={{Calculation of critical layer thickness versus lattice mismatch for $\text{Ge}_{x}\text{Si}_{1-x}$/Si strained-layer heterostructures}},
  author={People, R and Bean, JC},
  journal={Applied Physics Letters},
  volume={47},
  number={3},
  pages={322--324},
  year={1985},
  publisher={AIP Publishing}
}

@Misc{LandoltBornstein2001:sm_lbs_978-3-540-31355-7_212,
editor="Madelung, O.
and R{\"o}ssler, U.
and Schulz, M.",
title={{Silicon (Si) elastic moduli of Si-I: Datasheet from Landolt-B{\"o}rnstein - Group III Condensed Matter {\textperiodcentered} Volume 41A1$\alpha$: ``Group IV Elements, IV-IV and III-V Compounds. Part a - Lattice Properties'' in SpringerMaterials (https://doi.org/10.1007/10551045{\_}212)"}},
publisher={Springer-Verlag Berlin Heidelberg},
note="Copyright 2001 Springer-Verlag Berlin Heidelberg",
doi="10.1007/10551045_212",
url="https://materials.springer.com/lb/docs/sm_lbs_978-3-540-31355-7_212"
}
\end{document}